# Information estimations and analysis of structures

Alexander Shaydurov
McGill University, Montreal, Canada
Email: shaydurov@ieee.org

*Abstract*- **In this paper are described the results of the information analysis of structures represented by graphs. The obtained information estimation (IE) are based on an entropy measure of C. Shannon. The orthogonal characteristics of the graphs are taken into account: vertexes and contours (paths for the graphs of the type "tree"). Obtained IE is univalent both for the non-isomorphic and for the isomorphic graphs, algorithmically, it is asymptotically steady and has vector character. These IE can be used for the solution of the problems ranking of structures by the preference, the evaluation of the structurization of subject area, the solution of the problems of structural optimization.
Information estimations and method of the information analysis of structures it can be used in many fields of knowledge (Electrical Systems and Circuit, Image recognition, Computer technology, Databases and Bases of knowledge, Organic chemistry, Biology and others) and it can be base for the structure calculus.**

## I. Introduction

Analysis and design of the technical systems are connected with the development of the mathematical simulation and the improvement of the methods of evaluating the systems. In our opinion, the state of systems must contain not only evaluations of energy, material or economic characteristics, but also information parameters, so that the resulting state of system was represented in the material, energy and information spaces.

Extremely important is the information component of state, which determines the information potential of system.

By the common determination of system is set-theoretical, that presents any system the set of the diverse elements, united under various relations and connections, which form integral unity and intended for the purposeful activity [8,11]. The essential part of any system is its structure understood as steady it ordering, the variety of elements, with the specific (not random) connections between the forming structure parts. Structure is the invariant of system, determines its integrity and integral qualities, which are in the totality of elements.

At the present time there are some approaches to the determination of the information: combinatory, probabilistic, algorithmic [5,7].

Measure of a variety of elements, connections, relations, uncertainty and heterogeneity of their distribution in space - structures, it is expedient to consider as topological quantity of the information [1,10].

In the mathematical models the structure can be different methods, including and with the aid of the graphs, which are been the limit of the abstracting of the structure of system.

It should be noted that in the determination of system there are not only sets, but also they are indicated the mathematical operations above their elements. Frequently the elements of these sets are subsystems, i.e., they can be considered as the objects of categories [2,9]. Various sets of connections between the objects are considered as the morphisms, whose properties determine properties, variety and complexity of elements. A similar idea makes possible to use for purposes of simulation and design of technical systems the categories theory, giving exclusive value for the structures of the systems being investigated.

Now information properties of systems are not practically investigated. In particular, is absent the mathematical methods for obtaining the IE of the models of systems, represented by graphs, which make it possible to compare of different structures, to regulate structures on the preference.

Obviously, set of information parameters can be represented in the form of two subsets: relative to the stable subset of structural information parameters and more dynamic subset of functional information parameters.

So for example, the parameters of quality of electric power relate to the information parameters, intended to decrease the uncertainty of the regimes of the work of electrical customers (entropy of environment) and they can be classified as the dynamic parameters of the functional informativeness Power Electrical Systems (PES). In also the time the IE of structure PES or structures of electrical networks PES are static information characteristics.

The tasks of the information analysis of structures are actual in many field of knowledge, where the concepts are structured, they are connected with different relations and they can be represented by the graphs of different nature.

## II. The information estomationsts

The characteristics considered in the IE, must contain the basic topological properties of graphs, which can be expressed in the form of two classes: the characteristics of the complexity and metric's properties. As the measure of the complexity can be used the concepts of

"concentration", "branching", "valency", a quantity of connections or a quantity of entering this element components. Metric properties can be the measures of the "distance" in considered space, the degree of the closeness of elements.

The basic topological property of the graphs is connected with the incidence and contour matrices [4,6]:

$$M * N^t = 0, \quad (1)$$

where **M** is the incidence matrix of the graph with the crossed out basic vertex (BN); **N** is the contour matrix.

Presenting each matrix's row **M** (vertex) and **N** (contour) as vectors we obtain the condition of the orthogonality of these vectors, and generalized the condition of the orthogonality of vertex's and contour's spaces.

Therefore any universal IE, in the first place, must consider as the vertex's and contour's properties of the graph; in the second place, to have vector nature, since these properties are orthogonal.

*A. Vertex's components of information estimation*

As the estimation of the complexity were used the degrees of vertexes of the graph. For any connected graph the following formula is carried out [4,6]:

$$\sum_{i=1}^{K} \rho_i = 2L, \quad (2)$$

where $\rho_i$ is the degree of i-vertex on the graph; L is a quantity of branches on the graph; K is a quantity of vertexes.

The degrees of the vertexes of graph can be presented as sequences of the partition of integer 2L on K summand by different ways. The partitions must be graphic [6], i.e., with the aid of these partitions it is possible to form the tree. The conditions of existence any "graphic" form of the partitions take the form [6]:

$$\rho_i > 0, \forall i \in K; \quad L = K - 1. \quad (3)$$

Consequently, the IE of the graphs, which considers only the degrees of the vertexs is not univalent, since are possible two or more nonisomorphic structures, which have the identical partitions.

For obtaining the univalent estimations it is necessary to consider the additional characteristics of the graph.

Let us note that the graphs, which have identical partitions, are characterized by compactness, extent, branching, i.e., by metric properties.

We define center or bicenter of the graph as a vertex or the pair of adjacent vertexes (bicenter) with minimum eccentricity [3]. In the case of the bicenter, for the center of the graph starts that of the vertexes of bicenter, which is characterized by the minimum sum of the distances to the remaining vertexes of the graph. As the metric property of the vertex let us accept the sum of eccentricity of the center of graph and distance from the center to this vertex - the "remoteness" of the vertex.

During the analysis of unmarked, nonisomorphic graphs, the use of the proposed metric property together with the degrees of the vertexes of graph does make it possible to obtain univalent vertex's entropy invariants:

$$H_1 = -\sum_{i=1}^{K} \frac{\rho_i}{2L} \log_2 \frac{\rho_i}{2L} \sum_{i=1}^{K} \frac{\varepsilon_c + r_i}{\sum_{i=1}^{K}(\varepsilon_i + r_i)} \log_2 \frac{\varepsilon_c + r_i}{\sum_{i=1}^{K}(\varepsilon_i + r_i)}$$

, (4)

where $\rho_i$ is the degree of i-vertex; L is a quantity of the branches; $\varepsilon_c$ is eccentricity of the center; $r_i$ is the distance between i-vertex and center on the graph.

However, for example, for the graphs of Electrical or Computers Networks are characteristic the marked graphs, for which important is the functional destination of each vertex, in particular, it is marked the Base Note (BN), which is the source of power, the estimation (4) is converted:

$$H_1 = \sum_{i=1}^{K} \frac{\rho_i}{2L} \log_2 \frac{\rho_i}{2L} - \sum_{i=1}^{K} \frac{\varepsilon_b + r_i}{\sum_{i=1}^{K}(\varepsilon_b + r_i)} \log_2 \frac{\varepsilon_b + r_i}{\sum_{i=1}^{K}(\varepsilon_b + r_i)},$$

(5)

where $\varepsilon_b$ is eccentricity of selected BN on the graph.

During the determination of the distances on the graph was used the algorithm, developed and realized in software [10].

The obtained IE are univalent and are suitable for the practical use in the vertex's space of the graphs.

*B. Contour's components of information estimation*

The contour's space of the graph is the totality of the contours, represented by contour matrix - **N**. It is possible to note that the complexity of a contour influence several factors, one of which are complexities, entering this contour of the vertexes, i.e., the degrees of the vertexs.

Furthermore, the complexity of the contour can be expressed through a quantity of the branches are entered this contour. For **i**-contour this is the sum of the absolute values of **i**-row elements of the matrix **N** - $C_i$.

Metric properties in the contour's space they are expressed as the concept of "closeness" (distance) of contours, with the use of a frequency of branches - a quantity of the entrance of **j**-branch in the contours of the graph. By contour matrix the frequency of **j**-branch is expressed by the sum of the absolute values of **j**-column elements of the matrix **N** - $F^i$.

For the possibility of analysis in the contour's space of graphs of the type "tree" is introduced the concept of an "open contour"- a path, which is the sequence of the branches from the center (or the selected BN) of the graph to one of the terminal vertexs. The "open contours" are the paths of the graph are represented by the matrix of the paths - **W**, a quantity of rows of which is equal to the number of terminal vertexs, and a quantity of columns are equal to a quantity of the branches of the graph.

Are possible situations, then in the graph there are terminal vertexs ("open contour") and the closed contours. In this case they are calculated separately the matrix of the paths - **W** and the contour matrix - **N**. In the calculation IE participates the united matrix – **U**.

Contour's components of the IE, which considers the complexity of contours and thIEr "closeness" will take the form:

$$H_2 = -\sum_{i=1}^{P}\sum_{j=1}^{M_i} \frac{\rho_j}{\sum_{j=1}^{M_i}\rho_j} \log_2 \frac{\rho_j}{\sum_{j=1}^{M_i}\rho_j} - \sum_{i=1}^{P} \frac{C_i}{\sum_{i=1}^{P} C_i} \log_2 \frac{C_i}{\sum_{i=1}^{P} C_i} - \sum_{j=1}^{L} \frac{F^j}{\sum_{j=1}^{L} F^j} \log_2 \frac{F^j}{\sum_{j=1}^{L} F^j}, \quad (6)$$

where P is a quantity of contours and/or paths on the graph; $M_i$ is a quantity of vertexes in i-contour (path); $\rho_j$ is the degree of j-vertex, entering the i- contour (path); $C_i$ is a quantity of branches in i - contour and/or path; $F^j$ is the frequency of j - branch.

It should be noted that in the topological sense, the contour is not only dual element with respect to the vertex, but also the more complex information structural element.

The IE constructed in the orthogonal by the vertex's and contour's spaces can be used separately depending on the formulation of stated problem of structural analysis.

In the general case the resulting estimation has vector character:

$$\vec{H} = H_1 * \vec{i} + H_2 * \vec{j}, \quad (7)$$

where $\vec{i}$, $\vec{j}$ are the unit vectors of selected coordinate system.

Thus, the IE of the complexity of structures, being represented as graphs are the vector functional:

$$\vec{H} = \vec{F}(\vec{\rho}, \vec{\varepsilon}, \vec{r}, \vec{C_*}, \vec{F^*}), \quad (8)$$

where $\vec{\rho}$ is the vector of the degrees of vertexs; $\vec{\varepsilon}$ is the vector of the eccentricities of vertexs; $\vec{r}$ is the vector of the distances between the vertexs and the center of the graph; $\vec{C_*}$ is the vector of the estimations of the complexity of contours (a quantity of branches in the contours); $\vec{F^*}$ is the vector of the estimations of the "closeness" of contours.

Since the IE is the measure of the topological variety, which appears (created) during the design of the system, then it is expedient to consider it as the measure of the structurization, the estimation of the information potential of the system.

*C. Test results*

For the IE of structures was developed special software and calculations for the group of the graphs were carried out.

1.We investigate the IE on the complete group of nonisomorphic graphs - $G^8$. At 8 vertexes it is possible to construct 23 nonisomorphic trees, having 11 different partitions [6].

From the analysis of the structures examined it is possible to assume that proposed IE makes it possible to univalently identify nonisomorphic graphs and it is estimation unique up to isomorphism (Fig.1).

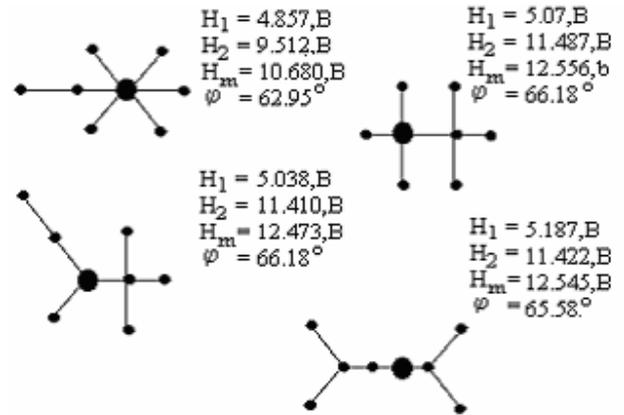

Fig 1. Information estimations for nonisomorphic graphs.

H1-vertex's component of IE; H2- contour's component; Hm-amplitude IE; φ-phase IE.

2. Analysis of the marked, isomorphic graphs and possible change the IE with different positions BN relative to the center of the graph. Smallest IE corresponding to the case, when BN coincides with the center of the graph. This conclusion should be considered, for example, during the design of Electrical or Computer's Networks. The Electrical Networks must be developed in such a way that BN as far as possible would coincide with the center of the Network (Fig.2).

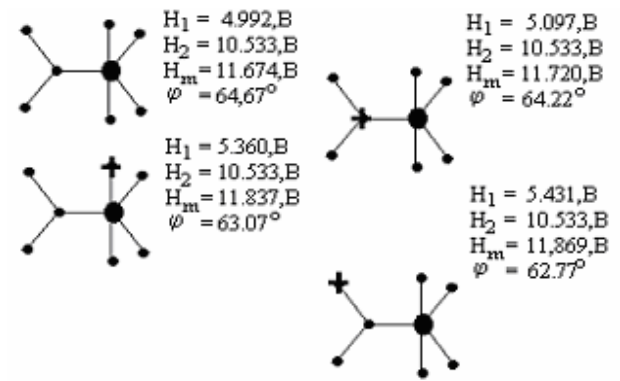

Fig 2. Influence a position of base vertex on the IE. Base Vertex marked + ; center graph – ●.

*D. Optimization*

In connection with the orthogonality of the vertex's and contour's spaces we will examine the extremum properties the IE separate. Furthermore, since the vertex's and contour's components of the IE are additive functions, then their extremum properties can be analyzed separately for each of the components.

1. The vertex's component of the IE (4,5) can be represented in the form:

$$H_1 = H_{11} + H_{12}, \quad (11)$$

where $H_{11}$ is the IE of the complexity of vertexs; $H_{12}$ is the information metric estimation.

- The component of the complexity of vertexes:

$$H_{11} = -\sum_{i=1}^{K} \frac{\rho_i}{2L} \log_2 \frac{\rho_i}{2L}. \quad (12)$$

After obvious conversions, we obtain:

$$H_{11} = \log_2 2L - \frac{1}{2L} \sum_{i=1}^{K} \rho_i \log_2 \rho_i. \quad (13)$$

1.1. If the limitation for the optimization task of the objective function (13) is the condition of connectedness (2), then Lagrangian will take the form:

$$\Phi(\overrightarrow{\rho, \lambda}) = \log_2 2L - \frac{1}{2L} \sum_{i=1}^{K} \rho_i \log_2 \rho_i + \lambda (2L - \sum_{i=1}^{K} \rho_i)$$

(14)

Determining the stationary points of Lagrangian, we obtain for the optimum estimations of degrees of vertexs and Lagrange's coefficient the following estimations:

$$\rho_i^{extr} = \frac{2L}{K}, \forall i \in K ; \quad \lambda^{extr} = -\frac{1}{2L} \log_2(eR). \quad (15)$$

The obtained optimum values $\rho_i^{extr}$ correspond to maximum value $H_{11}$.

If the considered structure is represented by complete graph, then the extreme estimation of degrees of vertexs $\rho_i^{extr} = K-1$, which corresponds to complete realization by each vertex of its potential connections.

1.2. The analysis of the condition of limited branching:

$$\rho_i \leq R, \forall i \in K, \quad (17)$$

where R is the maximum permissible branching the vertexs (maximum degree of the vertexs), we come to the conclusion that extreme value of the objective function will be during the maximum realization each vertex of their connections:

$$\rho_i = R, \quad \lambda_i^{extr} = -\frac{1}{KR} \log_2(eR), \forall i \in K, \quad (18)$$

where e = 2.71828.. is Naperian base.

Values $\lambda_i^{extr}$ determine the reaction of the objective function (13) to a change in the limitations R and make it possible to find those limitations, which it is expedient to change for an improvement in objective function.

Greatest estimation of $H_{11}$:

$$H_{11}^{extr} = \log_2 2L - \frac{1}{2L} \sum_{i=1}^{K} \frac{2L}{K} \log_2 \frac{2L}{K} = \log_2 K. \quad (19)$$

This estimation can be considered as the upper maximum estimation of $H_{11}$.

- Metric component:

$$H_{12} = -\sum_{i=1}^{K} \frac{\varepsilon_b + r_i}{\sum_{i=1}^{K}(\varepsilon_b + r_i)} \log_2 \frac{\varepsilon_b + r_i}{\sum_{i=1}^{K}(\varepsilon_b + r_i)}, \quad (20)$$

where $\varepsilon_b$ is eccentricity of BN or center of the graph; $r_i$ is distance from BN or center to i-vertex.

It is not difficult to prove that the maximum of the function (20) is determined by the condition:

$$\varepsilon_b + r_i \to const, \forall i \in K \quad (21)$$

and the extreme value of metric component:

$$H_{12}^{extr} = \log_2 K. \quad (22)$$

1.3. When the condition of the limited remoteness is present:

$$r_i \leq d, \forall i \in K, \quad (23)$$

where d is the assigned maximum permissible diameter of the graph, estimation will be extreme under the condition:

$$r_i = d, \quad \lambda_i^{extr} = -\frac{(K-1)}{K^2} \frac{1}{(\varepsilon_b + d)} \log_2 \frac{e}{K}, \quad \forall i \in K$$

(24)

As a whole, for the vertex's component of the IE extreme asymptotic estimation is determined:

$$H_1^{extr} = 2\log_2 K. \quad (25)$$

In this case it is necessary that degrees of the vertexes and their distance from the center or BN would be as far as possible constants.

2. The contour's component of the IE (7) is represented in the form:

$$H_2 = H_{21} + H_{22} + H_{23}, \quad (26)$$

where $H_{21}$ is the IE of the complexity of vertexes, entering contours or paths on the graph; $H_{22}$ is the IE of the complexity of contours (paths); $H_{23}$ is the information metric estimation of the contours (paths).

- The component of the complexity of vertexes, entering the contour:

$$H_{21} = -\sum_{i=1}^{P} \sum_{j=1}^{M_i} \frac{\rho_j}{\sum_{j=1}^{M_i} \rho_j} \log_2 \frac{\rho_j}{\sum_{j=1}^{M_i} \rho_j}. \quad (27)$$

2.1. In the presence of limitations to the complexity of vertexs and thIEr quantity in the contours:

$$\rho_j \leq R, \forall j \in M_i; \quad M_i \leq M, \forall i \in P, \quad (28)$$

for obtaining $H_{21}^{extr}$ is necessary satisfaction of the conditions for the complete utilization of the corresponding resources R and M:

$$\rho_j = R, \forall j \in M_i; M_i = M; \lambda^{extr} = -\frac{(M-1)}{M^2}\frac{1}{R}\log_2\frac{e}{R}$$
$$, \forall i \in P. \quad (29)$$

The extreme value the IE of the complexity of vertexes, entering the contours:

$$H_{21}^{extr} = P \log_2 M, \quad (30)$$

where P is a quantity of contours on the graph (if the graph has terminal vertexs, then this is the estimation of the sum of the closed contours and paths); M is a quantity of vertexs in the contour (path).

- The component of the complexity of contours:

$$H_{22} = -\sum_{i=1}^{P}\frac{C_i}{\sum_{i=1}^{P}C_i}\log_2\frac{C_i}{\sum_{i=1}^{P}C_i}. \quad (31)$$

The extreme estimation is determined by the constancy of a quantity the branches in the contours (paths):

$$C_i \to const, \forall i \in P. \quad (32)$$

2.2. If the limitation is assigned:

$$C_i \leq C_{max}, \forall i \in P, \quad (33)$$

determining maximum permissible quantity of branches, entering each contour (path) of the graph, we obtain the extreme value during the maximum use of the available resource $C_{max}$:

$$C_i = C_{max}, \quad \lambda_i^{extr} = -\frac{(P-1)}{P^2}\frac{1}{C_{max}}\log_2\frac{e}{P}, \forall i \in P \quad (34)$$

The extreme estimation of this component:

$$H_{22}^{extr} = \log_2 P. \quad (35)$$

- Metric component

$$H_{23} = -\sum_{j=1}^{L}\frac{F^j}{\sum_{j=1}^{L}F^j}\log_2\frac{F^j}{\sum_{j=1}^{L}F^j}. \quad (36)$$

The extreme estimation is determined by a constant frequency of the branches:

$$F^j \to const, \forall j \in L. \quad (37)$$

2.3. If the limitation, which determines the maximum permissible frequency of branches is assigned:

$$F^j \leq F_{max}, \forall j \in L, \quad (38)$$

then the extreme value of the objective function (36) is obtained during the complete utilization of maximum frequency of each branch:

$$F^j = F_{max}, \quad \lambda_j^{extr} = -\frac{(L-1)}{L^2}\frac{1}{F_{max}}\log_2\frac{e}{L}, \quad \forall j \in L$$
(39)

The extreme value of this component:

$$H_{23}^{extr} = \log_2 L. \quad (40)$$

As a whole, for the contour's component of the IE:

$$H_2 = P \log_2 M + \log_2 P + \log_2 L. \quad (41)$$

Comparing the asymptotic estimations of the vertex's and contour's components, it is not difficult to ascertain that with an increase in the complexity of the structure (K, L) the contour's components increases considerably more rapid than vertex's components.

CONCLUSION

1. Structural information is the objective property of the structure and determines its internal multiplicity and complexity. The obtained IE univalently reflect the topological variety of the structures and can be used for the ordering of the structures and estimation of the degree of structurization of subject areas.
2. Numerical values of the obtained information estimations make it possible to judge the complexity of the structure of system and for its correspondence to the decided tasks. Data of estimation, together with the "energy", "material", "financial" and by others they make it possible to more fully and thoroughly evaluate the of states of systems.
3. Structural information estimations are the initial stage of studies of the information analysis of the structures of systems and can be examined as the basis of the calculus of structures.